\documentstyle[12pt,amsbsy]{article}
\begin{document} 
\newcommand{\beq}{\begin{equation}}
\newcommand{\eeq}{\end{equation}}
\newcommand{\eq}[1]{(\ref{#1})}
\newcommand{\beqn}{\begin{eqnarray}}
\newcommand{\eeqn}{\end{eqnarray}}
\newcommand{\dst}{&\displaystyle}
\newcommand{\ba}{\bar \alpha}
\newcommand{\bb}{\bar \beta}
\newcommand{\fr}[2]{\frac{#1}{#2}}
\newcommand{\p}{\mbox{${\bf p}$}}
\newcommand{\q}{\mbox{${\bf q}$}}
\newcommand{\n}{\mbox{${\bf n}$}}
\newcommand{\bk}{\mbox{${\bf k}$}}
\newcommand{\bv}{\mbox{${\bf v}$}}
\newcommand{\s}{\mbox{${\bf s}$}}
\newcommand{\bS}{\mbox{${\bf S}$}}
\newcommand{\si}{\mbox{${\boldsymbol\sigma}$}}
\newcommand{\Si}{\mbox{${\boldsymbol\Sigma}$}}
\newcommand{\Nabla}{\mbox{${\boldsymbol\nabla}$}}
\newcommand{\vrho}{\mbox{${\boldsymbol\rho}$}}
\newcommand{\vzeta}{\mbox{${\boldsymbol\zeta}$}}
\newcommand{\vom}{\mbox{${\boldsymbol\omega}$}}
\newcommand{\bu}{\mbox{${\bf u}$}}
\newcommand{\bp}{\mbox{${\bf p}$}}
\newcommand{\E}{\mbox{${\bf E}$}}
\newcommand{\A}{\mbox{${\bf A}$}}
\newcommand{\B}{\mbox{${\bf B}$}}
\newcommand{\e}{\mbox{${\bf e}$}}
\newcommand{\f}{\mbox{${\bf f}$}}
\newcommand{\fib}{\mbox{${\boldsymbol\phi}$}}
\newcommand{\va}{\mbox{${\bf a}$}}
\newcommand{\ep}{\mbox{${\epsilon}$}}
\newcommand{\al}{\mbox{${\alpha}$}}

\vspace{1.5cm}

\begin{center} 
{\large \bf  Gauge Invariance and Canonical Variables}
\end{center}
\vspace{0.5cm}

\begin{center} 
I.B. Khriplovich\footnote{E-mail address: khriplovich@inp.nsk.su}
and A.I. Milstein\footnote{E-mail address: milstein@inp.nsk.su}\\
\end{center}

\vspace{0.5cm}

\begin{center}
Budker Institute of Nuclear Physics,\\
630090 Novosibirsk, Russia,\\
and Novosibirsk University, Novosibirsk, Russia
\end{center}

\vspace{1.5cm}

\begin{abstract}
We discuss some paradoxes arising due to the gauge-dependence of
canonical variables in mechanics.
\end{abstract}

\vspace{2.5cm}

{\bf 1}. Rather elementary problems discussed in this note originate 
partly from
tutorials on quantum mechanics at the Novosibirsk University, partly
from discussions on elementary particle physics and quantum field theory
with our colleagues. These problems turned out difficult not only for
undergraduates. To our surprise, they caused confusion even of some 
educated theorists. So, hopefully, a short note on the subject will be
useful, at least from the methodological point of view, so much the
more that we are not aware of any explicit discussion of the matter in
literature. 

Though the questions have arisen in quantum mechanics or even in more 
elevated subjects, they belong in essence to classical mechanics. Just
to classical mechanics we confine mainly in the present note.

\bigskip

{\bf 2}. Let us consider the simple problem of a charged particle in a
constant homogeneous magnetic field. Its Hamiltonian is well-known:
\beq\label{H}
H\,=\;{1 \over 2m}\,\left(\bp\,-\,{e \over c}\,\A \right)^2.
\eeq
It is also well-known that various gauges are possible for the vector
potential $\A$. With the magnetic field $\B$ directed along the $z$ 
axis, one can choose, for instance, 
\beq 
\A\,=\,B(0,\,x,\,0).
\eeq
In this gauge the Hamiltonian is independent of $y$, and therefore the
corresponding component $p_y$ of the canonical momentum is an integral 
of motion. However, one can choose equally well another gauge: 
\beq 
\A\,=\,B(-y,\,0,\,0).
\eeq 
Then it is the component $p_x$ of the canonical momentum which is 
conserved.

But how it comes that a component of $\bp$ transverse to the magnetic
field can be conserved, and that, moreover, the 
conserved component can be chosen at will? The obvious answer is that 
the canonical momentum 
$\bp$ is not a gauge-invariant quantity and therefore has no direct 
physical meaning. As to our visual picture of the transverse motion in
a magnetic field, it is not the canonical momentum $\bp$ which 
precesses and thus permanently changes its direction, but the velocity 
$$\bv\,=\,{1 \over m}\,\left(\bp\,-\,{e \over c}\,\A \right).$$
As distinct from the canonical momentum $\bp$, the velocity $\bv$ is a
gauge-invariant and physically uniquely defined quantity.

\bigskip

{\bf 3}. It is only natural that not only the space components $\bp$ of
the canonical momentum, but as well its time component, the 
Hamiltonian $H$, is gauge-dependent. It is the kinetic energy 
$H-eA_0$ which is gauge-invariant. 

As a rather striking manifestation of this fact, let us consider an 
example of a well-known physical system whose
energy is conserved, but the Hamiltonian can be time-dependent.
We mean the motion of a charged particle in a 
time-independent electric field $\E$, for instance, in the
Coulomb one. Let us choose here the gauge $A_0=0$. 
In it the vector potential becomes obviously
\[ \A\,=\,-c t \E, \]
so that now the Hamiltonian (\ref{H}) depends on time explicitly. 
Nevertheless, the energy of a particle in a time-independent
electromagnetic field is certainly conserved. Indeed, here the 
equations of motion become
\beq\label{vel}
\dot{\bf r}\,=\,\{H,\,{\bf r}\}\,=\,{1 \over m}\,(\bp\,+\,e\,t\,\E), 
\eeq
\beq\label{acc}
m\,\ddot{\bf r}\,=\,{d \over dt}\,(\bp\,+\,e\,t\,\E)
\,=\,\{H,\,\bp\,+\,e\,t\,\E \}\,=\,e\,\E 
\eeq
(we use the Poisson brackets $\{...,\,...\}$ in these classical 
equations). Since for a time-independent electric field its strength
can be always written as a gradient of a scalar function: 
$\E\,=\,-\mbox{\boldmath $\nabla$}\varphi$, equation (\ref{acc}) has
first integral
\[ {1 \over 2}\,m\,\dot{\bf r}^2\,+\,e\,\varphi\,=\,const\,, \] 
which is obviously nothing but the integral of energy. On the other 
hand, in virtue of equation (\ref{vel}), the Hamiltonian in the gauge
$A_0=0$ coincides in fact with the {\it kinetic} energy:
\[ H\,=\,{1 \over 2m}\,(\bp\,+\,e\,t\,\E)^2\,
=\,{1 \over 2}\,m\,\dot{\bf r}^2\,. \]
It looks quite natural: the kinetic energy $H-eA_0$, being 
gauge-invariant, should coincide with the Hamiltonian in the gauge
$A_0=0$.
 
At last, an obvious comment on the situation in quantum mechanics.
Though the Hamiltonian is not gauge-invariant, the Schr\"{o}dinger
equation 
is. Its gauge invariance is saved by the gauge transformation of the
wave function. In particular, in the gauge $A_0=0$ the time-dependence 
of the Hamiltonian results only in some extra time-dependent phase for
the wave function.

\begin{center}
***
\end{center}

We appreciate useful discussions with S.A. Rybak and V.V. Sokolov.
We are grateful to S.A. Rybak also for the 
advice to publish this note. The work was supported by 
by the Ministry of Education through grant No. 3N-224-98,
and by the Federal Program Integration-1998 through Project No. 274. 
 
\end{document}